\documentclass{article}

\usepackage{PRIMEarxiv}

\usepackage[utf8]{inputenc} 
\usepackage[T1]{fontenc}    
\usepackage{hyperref}       
\usepackage{url}            
\usepackage{booktabs}       
\usepackage{amsfonts}       
\usepackage{nicefrac}       
\usepackage{microtype}      
\usepackage{lipsum}
\usepackage{fancyhdr}       
\usepackage{graphicx}       
\graphicspath{{media/}}     
\usepackage{siunitx}
\usepackage{amsmath}
\usepackage{subcaption}
\usepackage{caption}
\usepackage{algorithm}
\usepackage{algpseudocode}
\usepackage{hyperref}
\title{DiffNMR3: Advancing NMR Resolution Beyond Instrumental Limits
}
\author{
  Sen Yan\thanks{\texttt{yansen0508@gmail.com}} \and Etienne Goffinet \and Fabrizio Gabellieri \and Ryan Young \and Lydia Gkoura \and Laurence Jennings \and Filippo Castiglione \and Thomas Launey \\
  Biotechnology Research Center,
  Technology Innovation Institute,
  Abu Dhabi,
  UAE\\
}

\begin{document}
\maketitle

\begin{abstract}
Nuclear Magnetic Resonance (NMR) spectroscopy is a crucial analytical technique used for molecular structure elucidation, with applications spanning chemistry, biology, materials science, and medicine. However, the frequency resolution of NMR spectra is limited by the "field strength" of the instrument. High-field NMR instruments provide high-resolution spectra but are prohibitively expensive, whereas lower-field instruments offer more accessible, but lower-resolution, results.
This paper introduces an AI-driven approach that not only enhances the frequency resolution of NMR spectra through super-resolution techniques but also provides multi-scale functionality. By leveraging a diffusion model, our method can reconstruct high-field spectra from low-field NMR data, offering flexibility in generating spectra at varying magnetic field strengths. These reconstructions are comparable to those obtained from high-field instruments, enabling finer spectral details and improving molecular characterization.
To date, our approach is one of the first to overcome the limitations of instrument field strength, achieving NMR super-resolution through AI. This cost-effective solution makes high-resolution analysis accessible to more researchers and industries, without the need for multimillion-dollar equipment.
\end{abstract}


\section{Introduction}

Nuclear Magnetic Resonance (NMR) spectroscopy is a powerful and widely utilized analytical technique for elucidating molecular structures \cite{marion2013introduction,gunther2013nmr}. By exploiting the magnetic properties of atomic nuclei, NMR provides rich information about molecular dynamics, chemical environments, and atomic connectivity. As a non-destructive technique, NMR has broad applications across various fields including chemistry, biology, materials science, and medicine. In chemistry, NMR is commonly used for identifying molecular structures, analyzing purity, and studying chemical reactions \cite{gunther2013nmr,duus2000carbohydrate}. In biology, it plays a critical role in the study of proteins, nucleic acids, and other biomolecules, often aiding in understanding complex processes \cite{jardetzky2013nmr}. NMR’s applications also extend to material science, where it helps characterize polymers, solid-state materials, and surfaces \cite{haouas2016recent}. Furthermore, in medicine, NMR principles are foundational for Magnetic Resonance Imaging (MRI), a key diagnostic tool in clinical settings \cite{brown2011mri}.

Despite its versatility, the performance of NMR spectroscopy, especially in terms of frequency resolution, is closely tied to the field strength of the instrument used \cite{becker1999high}. Frequency resolution refers to the ability of the instrument to resolve closely spaced resonance peaks, which is crucial for accurately identifying and characterizing molecules. Higher field strengths provide better resolution, allowing researchers to distinguish between subtle variations in chemical environments and obtain finer details about molecular structures. For example, high-field NMR instruments offer exceptional spectral clarity, resolving complex overlapping signals. However, such instruments come at a significant cost, often ranging from millions to tens of millions of dollars, making them inaccessible to many research institutions and industries \footnote{https://www.brucker.com}. 

Conversely, lower-field NMR instruments are more affordable, costing hundreds of thousands of dollars, but the limitation is lower frequency resolution \cite{blumich2019low}. The broader, less distinct peaks in low-field spectra can obscure important molecular details, making it difficult to fully analyze complex samples. This limitation in frequency resolution is a significant challenge in NMR spectroscopy, as it restricts the ability of researchers to extract accurate and detailed information from their samples, particularly when working with complex mixtures, large biomolecules, or subtle chemical environments.

This paper primarily focuses on addressing the limitation of frequency resolution, which is a key factor that affects the quality and interpretability of NMR spectra. 
To overcome this obstacle, we introduce a novel AI-driven approach that aims to overcome the limitations of instrument field strength by enhancing the frequency resolution of NMR spectra through \textbf{super-resolution} techniques \cite{park2003super,moser2024diffusion}. Super-resolution, in the context of NMR, refers to the process of improving spectral resolution beyond the native capabilities of the instrument. Our method leverages the diffusion model \cite{ddpm}, a type of AI model, to reconstruct high-field NMR spectra from low-field data. This AI model can generate enhanced spectra that reveal finer details typically hindered in low-field experiments by learning the relationships between low-field spectra and the corresponding high-field spectra.

Moreover, our approach offers \textbf{multi-scale functionality}, allowing for flexible reconstructions at varying magnetic field strengths. This means that the AI model is not limited to upscaling spectra to a fixed higher-field result (e.g., from 400 MHz to 900 MHz), but can generate spectra corresponding to intermediate field strengths as well. 
This flexibility is particularly valuable in tailoring the resolution to the specific needs of different experimental setups or research questions. Such reconstructions are not just approximate, but are \emph{comparable} to those obtained from high-field instruments, making the method a potential substitute for expensive high-field equipment. 

The significance of this work lies in its ability to circumvent the limitations of NMR instrument field strength, providing an alternative to expensive high-field NMR instruments. By providing a cost-effective solution that achieves multi-scale super-resolution without requiring costly hardware upgrades, this method makes high-field NMR data more accessible.
Thus, researchers and industries that previously relied on expensive equipment for high-resolution spectral analysis can now achieve similar outcomes using more affordable, lower-field instruments.

\section{Related work}
\subsection{Diffusion model for the super-resolution task.}
\begin{figure*}[tb]
    \centering
    \includegraphics[width=1\linewidth]{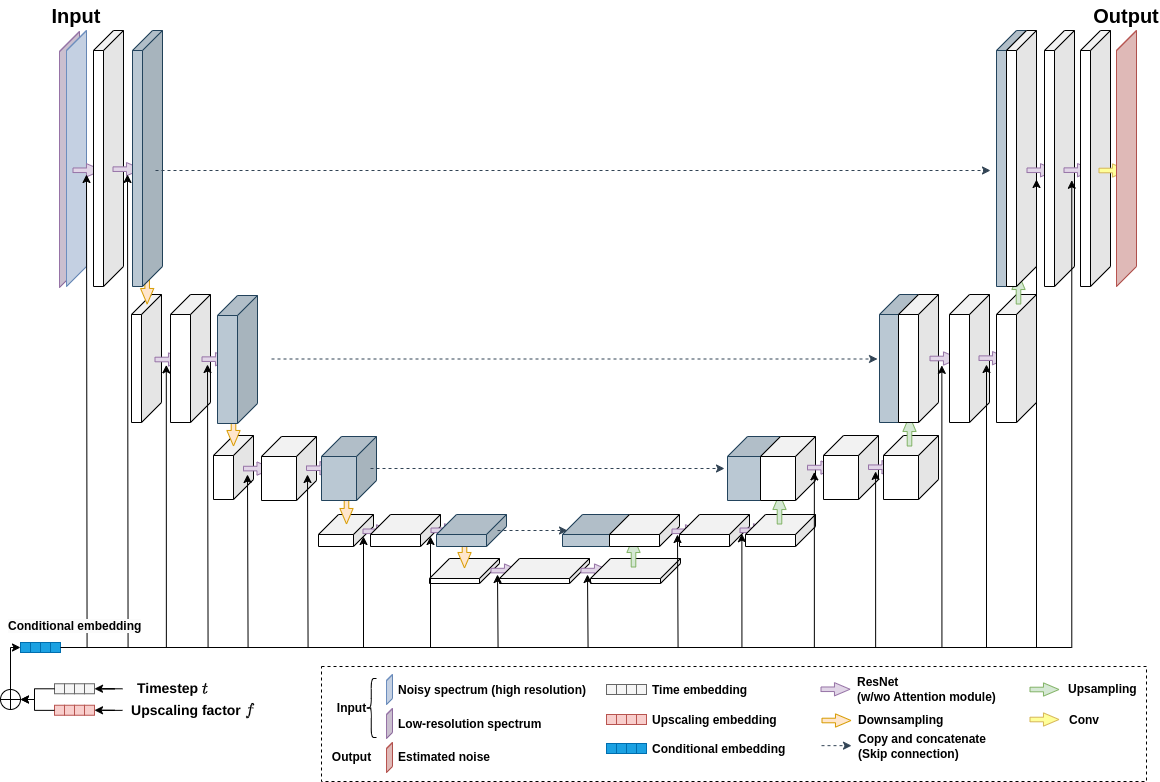}
    \caption{The backbone of our MSSR approach. The UNet architecture \cite{unet,AttentionUnet} is an encoder-decoder structure. 
    The model consists of a series of blocks for downsampling and upsampling, which form a U-shape.  In our MSSR approach, the noisy high-resolution spectrum and the low-resolution spectrum are first concatenated and fed to the first layer of the model and then forwarded sequentially through different blocks until the final outputs. The upscaling factor and timestep embeddings are summed and inputted to each block independently.}
    \label{fig:unet}
\end{figure*}

The explosion of interest in generative models came in the 2010s, as deep learning techniques became more mature. 
Generative models learn the underlying patterns and structures of the data, enabling them to produce novel instances. These models can generate a wide range of data types, such as images \cite{ganimation,yan2023combining,imga}, text \cite{bert,touvron2023llama}, and audio \cite{van2016wavenet,diffusion-audio}, and are capable of tasks like data augmentation \cite{sr3,srdiff}, content creation \cite{touvron2023llama}, and style transfer \cite{karras2019style}.
This period saw the development of key generative models such as Generative Adversarial Networks (GANs) \cite{gan}, Variational Autoencoders (VAEs) \cite{vae}, and more recently Diffusion models \cite{ddpm}.

Diffusion models are probabilistic models designed to learn a data distribution \cite{ddpm}. 
They are widely used in computer science, majorly in computer vision and visual-language models \cite{clip}.
The model operates by gradually adding noise (generally, Gaussian noise) to data in the forward process (called diffusion process) and then learning to reverse this noise through a backward process (called denoising process) to recover the original data distribution.
From 2020, the diffusion model family has gained attention due to its effectiveness in generating high-quality samples, initially in the context of image generation such as Repaint \cite{repaint}, then audio generation such as Diffwave \cite{diffusion-audio}, and more recently image-text generation such as Stable diffusion \cite{ldm}, GLIDE\cite{glide}, and Dall-E 2 \cite{dalle2}.

The backward process, which is the core of the diffusion model, aims to denoise the corrupted data in an iterative process. 
This reverse diffusion process in diffusion models generally relies on deep neural networks. As shown in Fig.~\ref{fig:unet}, UNet, a well-established architecture originally designed for biomedical image segmentation is widely used in diffusion models due to its multi-scale nature that facilitates the integration of fine and coarse features \cite{unet}. 
The UNet or its variants \cite{unet,AttentionUnet} is regarded as the backbone of the diffusion model.

Diffusion models are in principle capable of modeling
conditional distributions \cite{ldm,glide,dalle2}.
This can be implemented with a conditional UNet and paves the way to control the denoising process through auxiliary information (i.e., conditioning inputs) such as class labels \cite{dhariwal2021diffusion}, low-resolution data \cite{saharia2022image} and texts \cite{glide,dalle2}.
By conditioning on this information, the model can generate outputs that adhere to desired constraints or specifications.

In the context of super-resolution, the conditioning input typically consists of a low-resolution version of the target data. The diffusion model learns to enhance the spatial resolution by iteratively refining the details while preserving the overall structure and content of the input. This can be achieved by concatenating or integrating the low-resolution input with the noise-corrupted data at each step of the reverse diffusion process, or by integrating the auxiliary information into the model through additional channels or attention mechanisms \cite{vaswani2017attention}. For more information, please refer to the UNet with attention modules \cite{AttentionUnet}.
In the super-resolution task, the diffusion model usually takes the following inputs during the denoising process:
\begin{itemize}
    \item A low-resolution data, which serves as the conditioning input. This is often a downsampled version of the target high-resolution data.
    \item The current time step \( t \) provides the information to the model on how much noise has been added to the original spectrum.
    \item Optional auxiliary conditions, such as a class label or other features describing specific output characteristics. These conditions help guide the model to generate outputs aligned with the desired specifications.
\end{itemize}



Most super-resolution techniques involve processing image data or multimodal data that includes images \cite{luo2023image,chen2024hierarchical}. SR3 \cite{sr3} and SRdiff \cite{srdiff} have achieved high-quality super-resolution by manipulating the pixel domain. 
However, unlike previous works that process the pixel-wise input, in this paper, we apply the conditional diffusion model in the frequency-based domain. Furthermore, our approach offers multi-scale functionality (see Section \ref{Condition mechanism}).

\subsection{NMR super-resolution}\label{nmr super-reso}

The application of AI models for NMR super-resolution tasks is still in its early stages. To the best of our knowledge, prior work in this area has primarily focused on improving resolution through data acquisition (i.e., experimental optimization) rather than post-acquisition processing.

Mulleti et al. \cite{mulleti2017super} applied finite-rate-of-innovation sampling (FRI) \cite{vetterli2002sampling} to achieve super-resolution in NMR spectroscopy. By reconstructing signals with fewer measurements, their approach accurately resolves overlapping or broadened peaks, enabling precise estimation of chemical shifts and enhancing the analysis of complex systems.

Wenchel et al. \cite{wenchel2024super} focused on optimizing the experimental process itself. Their method dynamically increases the number of scans over time to counteract signal decay, reducing peak linewidth and enhancing resolution. However, this approach is limited to improving resolution during the data acquisition process and does not address the enhancement of pre-existing low-resolution spectra.

\begin{figure*}[btp]
    \centering
    \includegraphics[width=\linewidth]{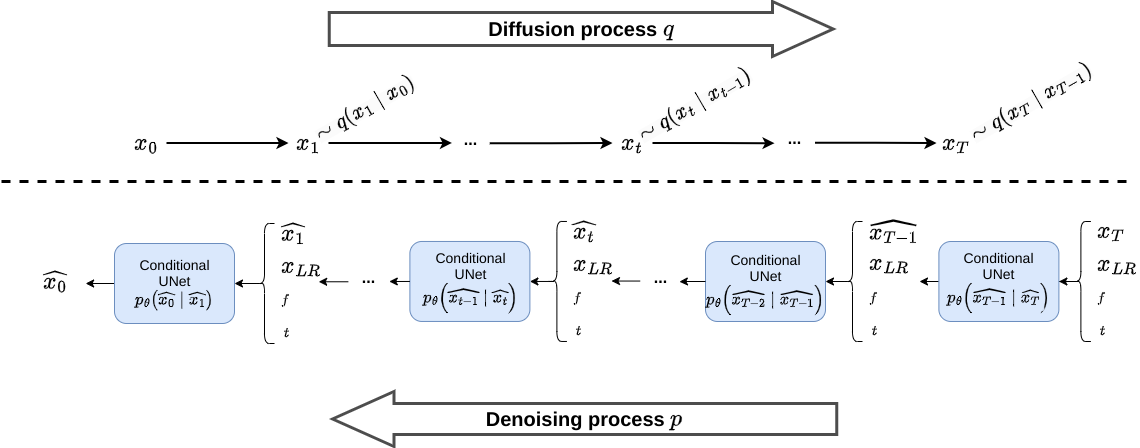}
    \caption{Our MSSR pipeline. In the diffusion process, the original NMR spectrum $x_0$ (high-resolution spectrum) is gradually corrupted by the Gaussian noise. After a total number of time steps $T$ in the diffusion process, the output is a noisy high-resolution spectrum denoted by $x_T$. $t$ represents the current time step.
    In the denoising process, an equally weighted sequence of UNets $p_{\theta}$, is trained to predict a denoised variant of their input $x_t$, where $x_t$ is a noisy version of the input $x_0$. $x_{LR}$ represents the low-resolution spectrum. $f$ is the upscaling factor representing the ratio of high resolution to low resolution.  At the end of the denoising process, the high-resolution spectrum denoted by $\widehat{x_0}$ is reconstructed. 
    The input of the UNet is the noisy spectrum $x_T$ at time step $T$ or the intermediate denoised spectrum (denoted by $\widehat{x_t}$) at time step $t$ where $t \in \{1,\dots,T-1\}$. The conditional inputs are the upscaling factor $f$, the time step $t$, and the low-resolution spectrum $x_{LR}$.
    Note that the architecture of the conditional UNet is detailed in Fig. \ref{fig:unet}.
    }
    \label{mssr}
\end{figure*}

Unlike the aforementioned works \cite{mulleti2017super, wenchel2024super}, we propose a novel AI-driven approach called the Multi-Scale Super-Resolution model (MSSR) shown in Fig.~\ref{fig:unet} and Fig.~\ref{mssr}. MSSR directly enhances the resolution of NMR spectra without requiring any improvements in the data acquisition process. By leveraging a conditional diffusion model, our method achieves super-resolution in post-acquisition processing, enabling the generation of high-resolution spectra from low-resolution spectra. This eliminates the reliance on costly high-field NMR instruments, providing an accessible and efficient alternative for improving spectral resolution.

\section{MSSR: Multi-Scale Super-Resolution pipeline}\label{DM}
As illustrated in Fig.\ref{mssr}, there are two processes in our MSSR pipeline: diffusion process and denoising process. In the diffusion process, the input is the original NMR (high-resolution) spectrum $x_0$. After adding noise to the input step by step, the final output is the noisy (high-resolution) spectrum $x_T$. 
In the denoising process, the noisy spectrum $x_T$ is fed to the conditional UNet (detailed in Fig. \ref{fig:unet}) with three conditioning inputs: the upscaling factor $f$, the time step $t$, and the low-resolution spectrum $x_{LR}$.
The conditional UNet gradually denoises the noise-corrupted spectrum. The high-resolution spectrum denoted by $\widehat{x_0}$ is reconstructed as the output. 

\subsection{Diffusion process} 
The diffusion process (i.e., the forward process) of the diffusion model is illustrated in the upper part of Fig.\ref{mssr}. The goal is to corrupt the spectrum by adding Gaussian noise step by step. 
A total number of time steps \(T\) is defined. In each time step \(t\), where $t \in \{1,\dots,T\}$), Gaussian noise is gradually added to the spectrum. The diffusion process $q$ can be described mathematically as:

\begin{equation}
    q(x_t | x_{t-1}) = \mathcal{N}(x_t; \sqrt{1 - \beta_t} x_{t-1}, \beta_t \mathbf{I})
\end{equation}
where \(x_0\) is the original spectrum;
\(x_t\) is the noisy spectrum at time step \(t\);
\(\beta_t\) represents the noise schedule, controlling the amount of noise added at step \(t\) (\(0 < \beta_t < 1\));
\(\mathbf{I}\) represents the Identity matrix.

\subsection{Denoising process}
The denoising process $p$ gradually removes noise to reconstruct the spectrum \(\widehat{x_0}\) (illustrated in the lower part of Fig.\ref{mssr}). 
It is modeled as a Markov process with learnable parameters:
\begin{equation}
    p_\theta(\widehat{x_{t-1}} | \widehat{x_t}) = \mathcal{N}(\widehat{x_{t-1}}; \mu_\theta(\widehat{x_{t}}, t), \Sigma_\theta(\widehat{x_{t}}, t)),
\end{equation}
where 
\(\mu_\theta(\widehat{x_{t}}, t)\): The learnable mean function parameterized by the UNet; 
\(\Sigma_\theta(\widehat{x_{t}}, t))\): The learnable covariance matrix parameterized by the UNet.

The reverse process iteratively denoises $x_T$ until $\widehat{x_0}$ is reconstructed.

\subsection{Condition mechanism}\label{Condition mechanism}
As shown in Fig. \ref{fig:unet}, the conditioning inputs are not directly fed to the conditional UNet. The projections are required to transfer the time step $t$ and the upscaling factor $f$ to a conditional embedding $z_{t,f}$. The low-resolution spectrum is concatenated to the input of the UNet (i.e., the noisy spectrum $x_T$ or the intermediate denoised spectrum $\widehat{x_t}$ with $t \in \{1,\dots,T-1\}$).

\textbf{Time embedding.} According to the literature \cite{ddpm,vaswani2017attention}, each time step $t$ is encoded using a sinusoidal positional encoding \cite{vaswani2017attention} to create a high-dimensional time embedding denoted by $\mathbf{m}_{t}$, with $t \xrightarrow{}\mathbf{m}_{t} \in \mathbb{R}^d$,
where \( d \) is the dimensionality of the embedding (pow of 2).

\textbf{Class condition.}
We discretize the upscaling factor $f$ by directly mapping each of the 
$n$ distinct upscaling factors to an integer index: $f \xrightarrow{}c \in \{0,1, \dots, n-1\}, \forall i\in[0, n-1]$ where $c$ is the class label.
For a given class labels $c$, we employ one-hot encoding \(\mathbf{h}_{c} \in \mathbb{R}^n\), where:
\[
\mathbf{h}_{c}[j] = 
\begin{cases} 
1, & \text{if } j = c, \\
0, & \text{otherwise.}
\end{cases}
\]
Then we generate the class embedding:
\[
\mathbf{w}_c = \mathbf{h}_c^\top \mathbf{W} \in \mathbb{R}^d.
\]
where $\mathbf{W} \in \mathbb{R}^{n \times d}$ is a learnable projection matrix; \(n\) represents the number of upscaling factors; \(d\) is the time embedding dimension.

Thus each upscaling factor $f$ is encoded as a unique class embedding \( \mathbf{w}_{\text{c}} \):
The time embedding and class embedding are added together to form the conditional embedding:

\begin{equation}
    \mathbf{z}_{t, f} = \mathbf{m}_{t} + \mathbf{w}_c
    \label{combined_emb}
\end{equation}

This conditional embedding $\mathbf{z}_{t, f}\in \mathbb{R}^d$ compressed the information of both the time step \( t \) and the upscaling factor \( f \).
This embedding is then integrated into the UNet’s convolutional blocks and the attention blocks, effectively providing the network with the information of the time step $t$ and the upscaling factor $f$. 

\textbf{Low-Resolution Spectrum.} 
The original spectrum (high-resolution) $x_0 \in \mathbb{R}^{h \times w}$ undergoes downscaling by expanding the peaks' widths based on a given upscaling factor $f$. Note that $h$ and $w$ represent the dimension of the spectrum.
Here, the input spectrum is convolved with the Gaussian window $g$.
Then the noise $\widetilde{\epsilon}$ is added to simulate the low-resolution spectrum. 
\begin{equation}
    x_{LR} = conv_g(x_0) + \widetilde{\epsilon}
\end{equation}
where
    \(x_{LR}\) is the resulting downscaled spectrum;
    \( g \) is the Gaussian kernel with standard deviation \(\sigma_{g} = \frac{1}{f}\);
    The Gaussian noise is denoted by $\widetilde{\epsilon} \in \mathbb{R}^{h \times w}$,
    \( \widetilde{\epsilon} \sim \mathcal{N}(0, \sigma_{LR}^2) \).

This low-resolution spectrum $x_{LR} \in \mathbb{R}^{h \times w}$ is concatenated with the noisy spectrum $x_T \in \mathbb{R}^{h \times w}$ or with the intermediate denoised spectrum $\widehat{x_t} \in \mathbb{R}^{h \times w}$ at different time steps. Therefore the conditional embedding $x_{cond}\in \mathbb{R}^{h \times w\times 2}$, which includes the noisy (high-field) spectrum and the corresponding low-field spectrum, is fed into the UNet. This allows the model to leverage the context of both spectra to reconstruct the original spectrum during the denoising process.

So far, the upscaling factor $f$, the time step $t$, and the downscaled spectrum $x_{LR}$ form the three conditioning inputs of the UNet.

\subsection{Training}
Firstly in the diffusion process, the noisy spectrum at time step $t$ (denoted by $x_t$) is generated by adding $t$ times of Gaussian noises from the original spectrum $x_0$, with $t$ uniformly sampled from $\{1,\dots,T\}$, where $T$ is the total number of time steps. 
Then in the denoising process, the UNet is trained to denoise the noisy spectrum $x_t$ step by step (from time step $t$ to 0).
The training objective is to minimize the Mean Reconstruction Error (MSE) loss $\mathcal{L}$ between the noise added to the original spectrum and the predicted noise.
\begin{equation}
    \mathcal{L} = \mathbb{E}_{x_0,  \epsilon \sim \mathcal{N}(0, 1),t,f} \left[ \left\| \epsilon - \epsilon_\theta(x_t,x_{LR}, t, f) \right\|_2^2 \right]
\end{equation}
Where $\epsilon$ is the noise added to the original spectrum; \( \epsilon_\theta(x_t,x_{LR}, t, f) \) is the noise predicted by the UNet.

\subsection{Inference}
For the inference, given a upscaling factor $f$ and a low-resolution spectrum $x_{LR}$, the model should reconstruct the spectrum $\widehat{x_0}$ from random noise sampled from the normal distribution.
That is to say, at time step $T$, the input of the conditional UNet is $x_T \sim \mathcal{N}(0, 1)$. The three conditioning inputs of the conditional UNet are unchanged.  The output of the UNet is controlled by the conditioning inputs. The final output is $\widehat{x_0}$ at time step $t=0$, with $\widehat{x_0} \approx x_0$.  

\section{Experiments and Results}
\subsection{Dataset and experiment configuration}
\textbf{Dataset.} Massive datasets containing hundreds of millions of images have been used in recent advances in diffusion models for image generation \cite{ldm,laion}.
In contrast, the availability of large-scale datasets for NMR protein spectra is extremely limited, with only a few public repositories accessible. 
This paper concentrates on enhancing the resolution of 2D NMR spectra.
The dataset used in this study is the 100-protein NMR spectra dataset (ARTINA) \cite{artina}, comprising 1329 spectra in 2D, 3D, and 4D formats. 
These spectra, derived from 100 proteins from real life, were sampled using NMR machines operating at
frequencies ranging from 600 to 950 MHz.
To expand the dataset, we included additional 2D spectra by projecting multi-dimensional (3D/4D) spectra into 2D representations. 
This technique, commonly used in NMR workflows for visualization purposes, increases the total
number of samples to over 3500. 
The process maintains the integrity of the original signals while significantly enriching the dataset for training.

We train our MSSR model on the ARTINA dataset with a train-validation-test ratio of 0.8, 0.1, and 0.1. This split ensures the test set's independence, minimizing the data leakage risk and guaranteeing robust performance evaluation.
All the data in ARTINA are proteins whose entries are listed in the Protein Data Bank \cite{pdb}.
The original spectra $x_0$ are normalized to $[-1,1]$ and resized to $x_0 \in \mathbb{R}^{256 \times 256}$.

\begin{table}[tbp]
    \centering
    \begin{tabular}{|c|*{5}{c|}}
    \hline
    \textbf{High | Low} & 400 & 500 & 600 & 700 & 800 \\
    \hline
    500 & 1.25 & -    & -    & -    & -    \\
    \hline
    600 & 1.50 & 1.20 & -    & -    & -    \\
    \hline
    700 & 1.75 & 1.40 & 1.17 & -    & -    \\
    \hline
    800 & 2.00 & 1.60 & 1.33 & 1.14 & -    \\
    \hline
    900 & 2.25 & 1.78 & 1.50 & 1.29 & 1.12 \\
    \hline
\end{tabular}

    \caption{Upscaling factor table. The upscaling factors are calculated as the ratio of the high-resolution frequency to the low-resolution frequency (\textit{High/Low}). 
        For example, the upscaling factor for \(500\,\text{MHz}/400\,\text{MHz}\) is \(1.25\). 
        Entries marked with '-' indicate that the calculation is not applicable since the low frequency should be smaller than the high frequency. 
        }
    \label{tab:reduced_factor}
\end{table}

\textbf{Upscaling factors.} Table \ref{tab:reduced_factor} provides upscaling factors for frequencies ranging from 400 MHz to 900 MHz, commonly used in NMR spectroscopy.  
Note that the upscaling factor for \(600\,\text{MHz}/400\,\text{MHz}\) and the upscaling factor for \(900\,\text{MHz}/400\,\text{MHz}\) are the same.
Thus there are totally $n=14$ upscaling factors ($n$ is defined in Section \ref{Condition mechanism}).

\textbf{Low-resolution spectrum.} Considering that the standard practice in NMR spectroscopy has been to train models on
simulated spectra \cite{zhan2024fast,qu2020accelerated,karunanithy2021fid,zheng2022fast},
we employ the simulated spectrum mentioned in Section \ref{Condition mechanism} as the low-resolution spectrum. Validated by the lab expert, we set $\sigma_{LR} = 0.01$.

\textbf{Training and Inference.} 
Considering the configuration of diffusion model-based approach \cite{repaint,ldm,glide,sr3}, we set the training time step $T=2000$.
Based on our empirical study in the Appendix (Fig. \ref{fig:metrics_vs_timesteps}), we set that the model infers 5 times with a fixed inference time step of $T=500$ to reconstruct spectra. Note that each inference produces a reconstructed spectrum.  
For each NMR super-resolution task, the final reconstruction of the NMR spectrum is the average of the reconstructed spectra from the 5 inferences.

\subsection{Baseline}
As illustrated in Fig. \ref{fig:unet} and Fig. \ref{mssr}, our MSSR approach is a unified model with multi-scale functionality. Indeed, this model can realize super-resolution from different low frequencies to different high frequencies, controlled by the 14 upscaling factors listed in Table \ref{tab:reduced_factor}.

To create a comprehensive baseline for comparison, we trained 14 different models, each tailored to one of the 14 upscaling factors. 
The difference between the 14 baseline models and our MSSR model is the conditioning input. For each baseline model, we eliminated the class embedding, so there is only time embedding $\mathbf{m}_{t}$ as the conditional embedding (see Section \ref{Condition mechanism}).
For the low-resolution spectrum, each model only creates only one group of the low-resolution spectra $x_{LR}$ by a fixed Gaussian kernel with standard deviation $\sigma_g=\frac{1}{f}$, where $f$ is the given upscaling factor. 
Since the upscaling factor is no longer the conditioning input, each model works for one super-resolution task (corresponding to an assigned upscaling factor).
Furthermore, the upscaling factor $f$ is removed from the loss function, $\mathcal{L} = \mathbb{E}_{x_0,  \epsilon \sim \mathcal{N}(0, 1),t} \left[ \left\| \epsilon - \epsilon_\theta(x_t,x_{LR}, t) \right\|_2^2 \right]$.


\begin{figure*}[btp]
    \centering
    \includegraphics[width=0.8\linewidth]{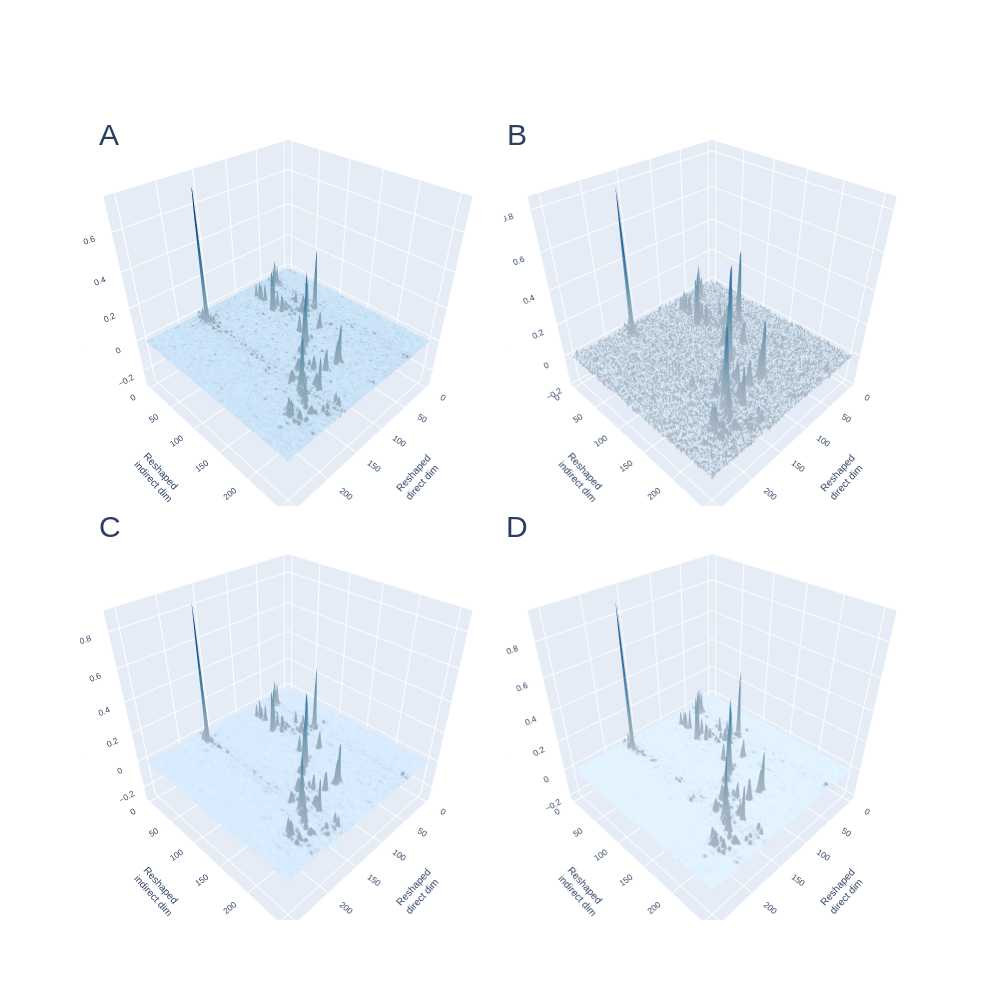}
    \caption{A: Original spectrum. B: Low-resolution spectrum (with upscaling factor $f=2$). C: \textbf{MSSR}. D: Baseline.}
    \label{results}
\end{figure*}

\subsection{Results}
We illustrate in Figure \ref{results}, the original spectrum of the compound 2KZV from the ARTINA dataset (2KZV, experiment: HCCHCOSY$@$ALI, nucleus: C, HC), the low-resolution spectrum, the reconstruction from MSSR and the reconstruction from the baseline.
Notably, both the baseline and our MSSR approach achieve impressive reconstruction quality; however, our approach excels in capturing finer details compared to the baseline. 
To facilitate a more nuanced evaluation of these methods, we have developed various metrics for comparison.

\begin{figure*}[btp]
    \centering
    \includegraphics[width=0.7\linewidth]{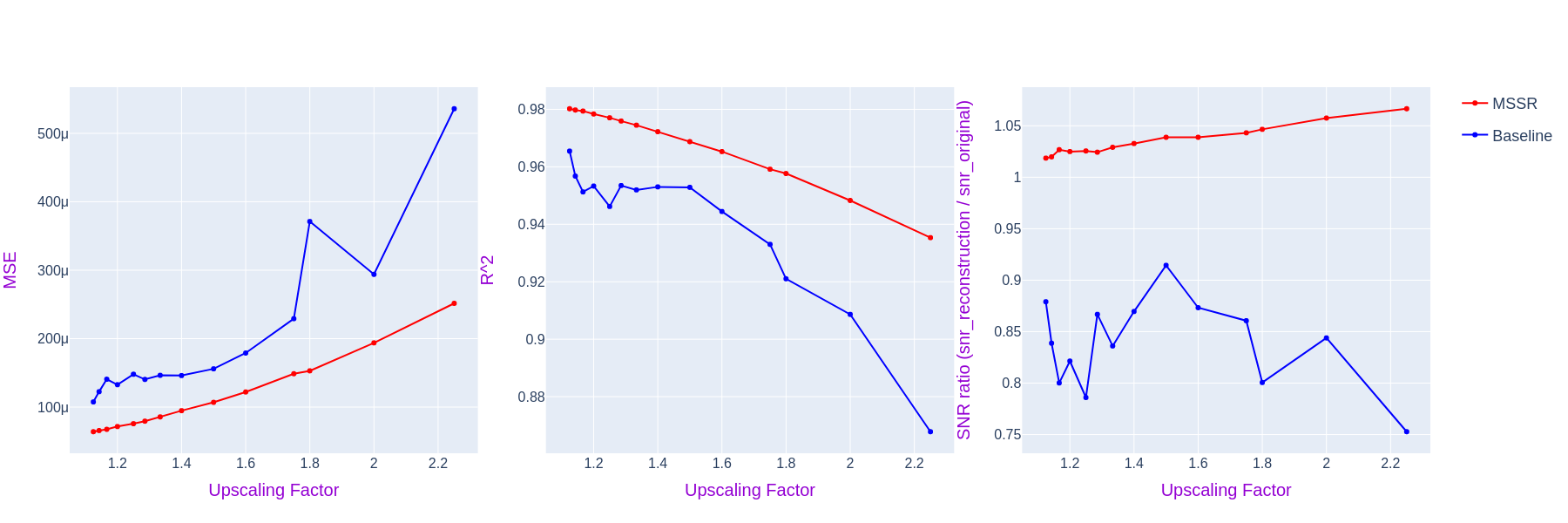}
    \caption{Global metrics. We investigate the Mean Squared Error (MSE) and the Coefficient of Determination ($R^2$). These metrics can measure the overall alignment and similarity between the original spectrum and the reconstructed spectrum. MSE: the lower the better. $R^2$ the higher the better.}
    \label{global}
\end{figure*}

\subsection{Metrics}
The evaluation approach balances both a global perspective, assessing the overall agreement between the original and reconstructed spectra, and a local perspective, focusing on the accuracy of individual spectral peaks, which are essential for compound characterization. 
Together, we select nine metrics to provide a comprehensive assessment of our model's performance.

\subsubsection{Global Metrics}
Global metrics measure the overall alignment and similarity between the reconstructed and original spectra. 
We select Mean Squared Error (MSE) and Coefficient of Determination ($R^2$).
MSE (ranging from 0 to infinity) quantifies the magnitude of prediction errors, ensuring accuracy in capturing
critical spectral features like peak intensities and positions. 
A lower MSE indicates higher accuracy in the reconstruction, as it shows that the overall deviation between the two spectra is minimized.
$R^2$ (ranging from 0 to 1) complements this by assessing how well the model
explains the variance in the data, providing a normalized measure of fit. An $R^2$ value close to 1 indicates that the reconstruction is highly representative of the original data, demonstrating a good fit.
Together, these metrics ensure a robust
evaluation of the method’s precision and reliability.



\subsubsection{Local Metrics}
Local metrics focus on evaluating the accuracy of specific spectral peaks, which are critical for identifying compound features. These metrics assess how well the AI model reproduces the precise details of the spectrum, such as the positions and intensities of the peaks, which are necessary for accurate interpretation.

We investigate the following peak-focused metrics: the hallucination ratio and the missed peak ratio, the peak MSE, peak $R^2$, the peak intensity difference, and the peak coordinate shift difference.

The hallucination ratio (ranging from 0 to 1) measures 
the proportion of peaks detected in the predicted spectrum that do not correspond to peaks in the original
spectrum (analogous to the False Detection Rate). 
These peaks are errors introduced by the model and can distort the interpretation of the spectrum. Minimizing this ratio is important to avoid introducing false features in the reconstructed data.

Differing from the hallucination ratio, the missed peak ratio (ranging from 0 to 1) measures the proportion of peaks in the original spectrum that do not appear in the reconstructed spectrum (analogous to the False Negative Rate). 
Reducing this ratio ensures that critical peaks are retained in the reconstruction.
Both the hallucination ratio and the missed peak ratio are calculated based on peaks identified
by a consistent expert system using the same parameters for all spectra, ensuring fair and unbiased comparisons.

The MSE of peaks (ranging from 0 to infinity) is similar to the global MSE but is focused specifically on the peaks of the spectra. It calculates the mean squared difference between the corresponding peaks in the original and reconstructed spectra. A lower value for this metric suggests that our MSSR model is accurately reconstructing the critical features of the spectrum, such as peak positions and heights.

The $R^2$ of peaks (ranging from 0 to 1) evaluates how well the positions and intensities of the peaks in the reconstructed spectrum match those in the original spectrum. A high $R^2$ value indicates that both the peak locations and their intensities are well-reconstructed, which is crucial for accurately representing the key spectral features.

The Peak Coordinates Shifting Difference (ranging from 0 to 1) measures the positional difference (shift) of corresponding peaks between the original and reconstructed spectra. Large shifts can mislead interpretations. Minimizing the peak coordinates shifting difference ensures that the reconstructed peaks are aligned correctly with the original ones.

\textbf{Peak Intensity Difference.} 
This Peak Intensity Difference (ranging from 0 to 1) compares the intensity (height) of matched peaks between the original and reconstructed spectra. Minimizing the peak intensity difference ensures that the relative abundance or concentration of compounds is accurately preserved in the reconstruction.

\subsection{Discussion}
\textbf{The global metrics} are illustrated in Figure \ref{global}.
MSSR consistently outperforms the Baseline with significantly lower MSE values. This indicates a smaller overall reconstruction error.
MSSR achieves higher $R^2$ values compared to the Baseline. As the upscaling factor increases, MSSR's $R^2$ approaches 1, whereas the Baseline shows slower improvement. This highlights MSSR's advantage in globally fitting the original spectra.

\begin{figure*}[btp]
    \centering
    \includegraphics[width=\linewidth]{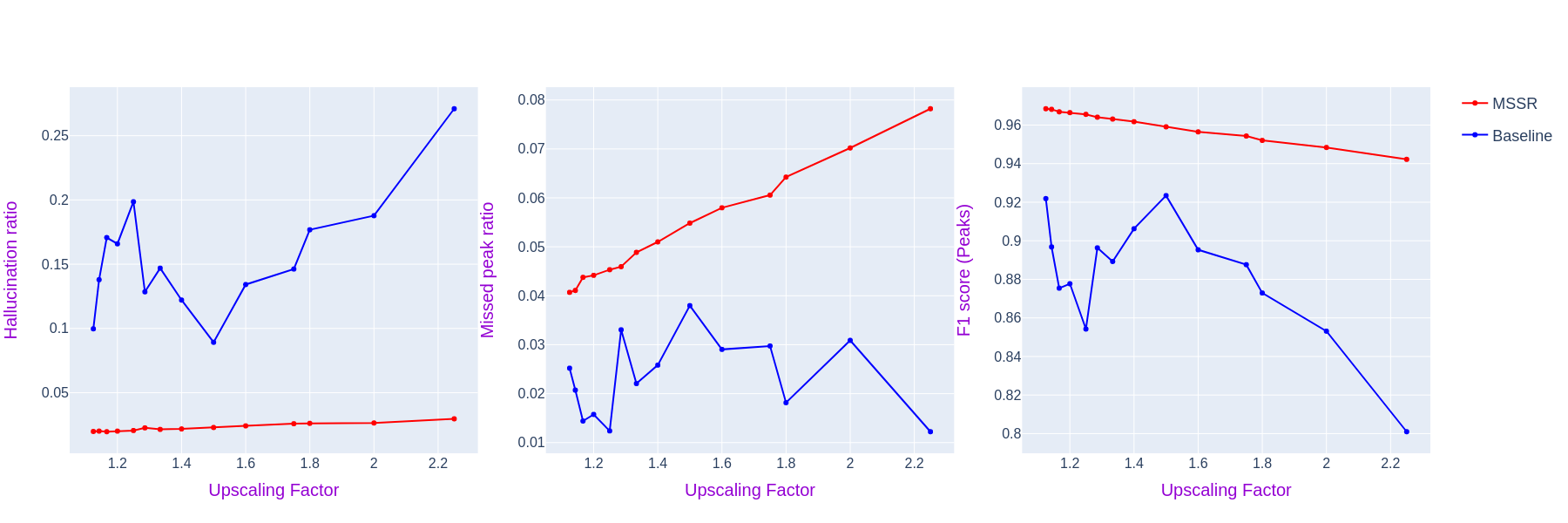}
    \caption{Local metrics. We list the peak-focused metrics since the peaks are critical for identifying compound features. For the $R^2$ of peaks: the higher the better, for the other metrics: the lower the better.}
    \label{local}
\end{figure*}

\textbf{Local metrics} are presented in Figure \ref{local}.
For the hallucination ratio, both our MSSR and the Baseline maintain a relatively low ratio (less than $3\%$). 
The MSSR curve is consistently lower than the Baseline, indicating that MSSR produces fewer artifacts. This is crucial for ensuring the authenticity of the generated spectra. 
For the missed peak ratio, the Baseline is unstable when the upscaling factor is smaller than 0.57. The difference between our MSSR and the Baseline becomes smaller as the upscaling factor increases. This suggests that MSSR captures true peaks more effectively across different scales.
For the peak MSE, MSSR achieves lower Peak MSE across all upscaling factors, particularly in the upscaling factors ranging from 0.57 to 0.89. This indicates MSSR reconstructs peak intensities with higher accuracy.
About $R^2$, MSSR consistently exhibits higher $R^2$ values compared to the Baseline. This demonstrates that MSSR excels in modeling the correlation between reconstructed and original peaks.
For the peak intensity difference, 
MSSR shows a significantly lower peak intensity difference compared to the Baseline. 
For the peak coordinate shift difference, the two curves are close. 
However, MSSR performs slightly better when the upscaling factors increase.

MSSR demonstrates superiority in reducing artifacts, capturing true peaks, minimizing errors, and improving correlation modeling. Its adaptability to multi-scale spectral characteristics is particularly evident under low upscaling factor conditions.
Globally, MSSR better preserves the overall structure and features of the original signal.
Overall, 
MSSR outperforms the Baseline across global and local metrics.

\section{Conclusion}
In this paper, we have introduced a novel AI-driven approach, called the Multi-Scale Super-Resolution model (MSSR), for enhancing the frequency resolution of NMR spectra using a conditional diffusion model to achieve multi-scale super-resolution. 
This method addresses one of the key limitations of NMR spectroscopy, i.e., frequency resolution, by reconstructing high-resolution spectra from low-resolution spectrum. Our approach enables researchers to obtain spectral detail comparable to that of high-field instruments without the associated cost, making high-quality NMR analysis more accessible to a broader audience.

The diffusion model used in this study provides flexible super-resolution capabilities, allowing the reconstruction of spectra at varying field strengths. This multi-scale functionality offers significant advantages in tailoring the resolution to specific experimental needs, making the method adaptable across different research and industrial contexts. 

The implications of this work can be far-reaching. By democratizing access to high-quality NMR data, this method has the potential to transform the way NMR is utilized in both academic and industrial settings. 
Researchers working with limited budgets can now achieve high-resolution results using affordable, lower-field NMR instruments.
By reducing reliance on expensive high-field NMR machines, this approach makes NMR spectroscopy more accessible, thus broadening the scope and potential of molecular analysis across scientific disciplines.

Further improvements to the model's accuracy and scalability could extend the range of applications, including more complex molecular systems or higher-dimensional or multimodal NMR data. Additionally, integrating our MSSR method with other techniques such as Non-Uniform Sampling (NUS) \cite{qu2020accelerated,zhan2024fast} may further optimize NMR analysis, opening up new possibilities for real-time, high-resolution molecular insights.


\begin{figure*}[tbp]
    \centering
    \includegraphics[width=\linewidth]{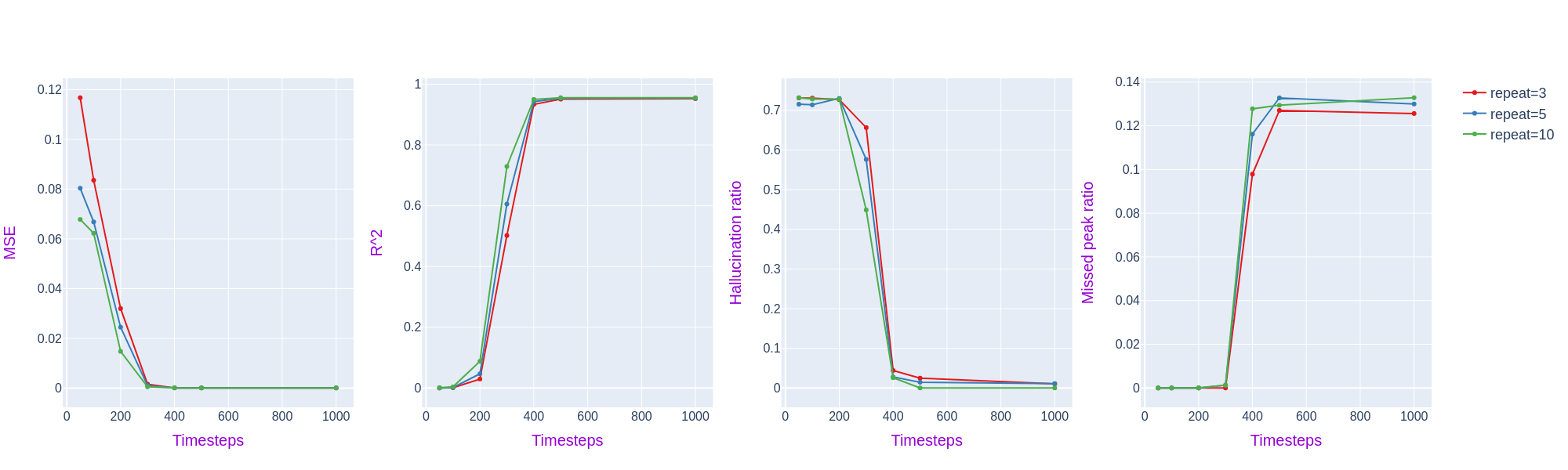}
    \caption{Different metrics plotted against the inference timesteps for different repeat values. Considering the consuming time for inference and the performance of the model, we set the inference time step to 500 and the inference repeat to 5.}
    \label{fig:metrics_vs_timesteps}
\end{figure*}
\section*{Appendix}\label{Appendix}
\textbf{Hyperparameters for inference.}
Fig. \ref{fig:metrics_vs_timesteps} illustrates the empirical study
related to the effect of time steps on various metrics (MSE, $R^2$, Hallucination ratio, Missed matching ratio) for different times of inferences (i.e., $repeat=3, repeat=5, repeat=10$). 
Increasing the number of time steps improves all metrics significantly up to a certain threshold (between 200 and 400 time steps), after which performance stabilizes.
Higher repeat settings (repeat=10) consistently lead to better results across all metrics.
Considering the configuration of the literature \cite{repaint,ldm,glide,sr3} and the trade-off between the inference time and performance, we set the inference time step $T$ to 500 and the inference repeat to 5.

\textbf{Peak Matching.} 
Peak matching identifies corresponding peaks between the reconstructed and original spectra. As shown in Algo.\ref{detect peaks}, we detect the local maximum to identify the peaks.

Once we have identified the peaks in both the original and reconstructed spectra, we calculate the Euclidean distance between each pair of peaks. By using a distance threshold, we can categorize the peaks as matched or unmatched, which allows us to determine the Hallucination Ratio and the Missed Peak Ratio.

\begin{algorithm}[hbt]
\caption{Peak Detection}
\label{detect peaks}
\begin{algorithmic}[1]
\Require Spectrum $x$, threshold factor $\alpha$, smoothing parameter $\sigma$, neighborhood size $n$
\Ensure Peak coordinates and amplitudes

\textbf{Step 1:} Apply Gaussian filter with standard deviation $\sigma$.

$x_{\text{smooth}} \gets \text{GaussianFilter}(x, \sigma)$

\textbf{Step 2:} Create a neighborhood matrix of size $n \times n$.

$N \gets \text{ones}(n, n)$

\textbf{Step 3:} Detect local maxima in $x_{\text{smooth}}$.

$L_{\text{max}} \gets (\text{maximum\_filter}(x_{\text{smooth}}, N) == x_{\text{smooth}})$

\textbf{Step 4:} Compute threshold based on mean and standard deviation of $x_{\text{smooth}}$.

$T \gets \text{mean}(x_{\text{smooth}}) + \alpha \cdot \text{std}(x_{\text{smooth}})$

\textbf{Step 5:} Identify peaks where local maxima exceed the threshold.

$P \gets \{(i, j) \mid L_{\text{max}}(i, j) \ \text{and} \ x_{\text{smooth}}(i, j) > T\}$

\textbf{Step 6:} Extract peak coordinates and amplitudes.

$\text{peak\_coords}, \text{peak\_amplitudes} \gets P, x[P]$

\Return $\text{peak\_coords}, \text{peak\_amplitudes}$

\end{algorithmic}
\end{algorithm}

\bibliographystyle{unsrt}  
\bibliography{references}

\end{document}